\documentclass{article}
\usepackage{spconf,amsmath,graphicx}
\usepackage{amsthm,amsmath,amssymb} 
\usepackage{mathrsfs}
\usepackage{lineno}
\usepackage{txfonts} 
\usepackage{enumitem}
\usepackage{multirow} 
\usepackage{caption} 
\usepackage{txfonts} 
\usepackage{array}
\usepackage{tabularx}
\usepackage{booktabs} 
\usepackage{longtable}
\usepackage{bm}
\usepackage[numbers,sort&compress]{natbib}

\title{A Long-Tail Friendly Representation Framework \\ for Artist and Music Similarity}
\name{Haoran Xiang$^{1,*}$\thanks{*The author conducted this work as an intern at ByteDance.}, Junyu Dai$^{2,**}$\thanks{**Corresponding Author.}, Xuchen Song$^2$, Furao Shen$^{1}$}
\address{$^1$National Key Laboratory for Novel Software Technology, Nanjing University, China \\
$^2$ByteDance}
\begin{document}
%
\maketitle
\begin{abstract}
The investigation of the similarity between artists and music is crucial in music retrieval and recommendation, and addressing the challenge of the long-tail phenomenon is increasingly important. This paper proposes a Long-Tail Friendly Representation Framework (LTFRF) that utilizes neural networks to model the similarity relationship. Our approach integrates music, user, metadata, and relationship data into a unified metric learning framework, and employs a meta-consistency relationship as a regular term to introduce the Multi-Relationship Loss. Compared to the Graph Neural Network (GNN), our proposed framework improves the representation performance in long-tail scenarios, which are characterized by sparse relationships between artists and music. We conduct experiments and analysis on the AllMusic dataset, and the results demonstrate that our framework provides a favorable generalization of artist and music representation. Specifically, on similar artist/music recommendation tasks, the LTFRF outperforms the baseline by 9.69\%/19.42\% in Hit Ratio@10, and in long-tail cases, the framework achieves 11.05\%/14.14\% higher than the baseline in Consistent@10.
\end{abstract}
\section{Introduction}
\label{sec:intro}

\begin{figure}[!t]
\centering\includegraphics[scale=0.35,trim=0 0 0 0]{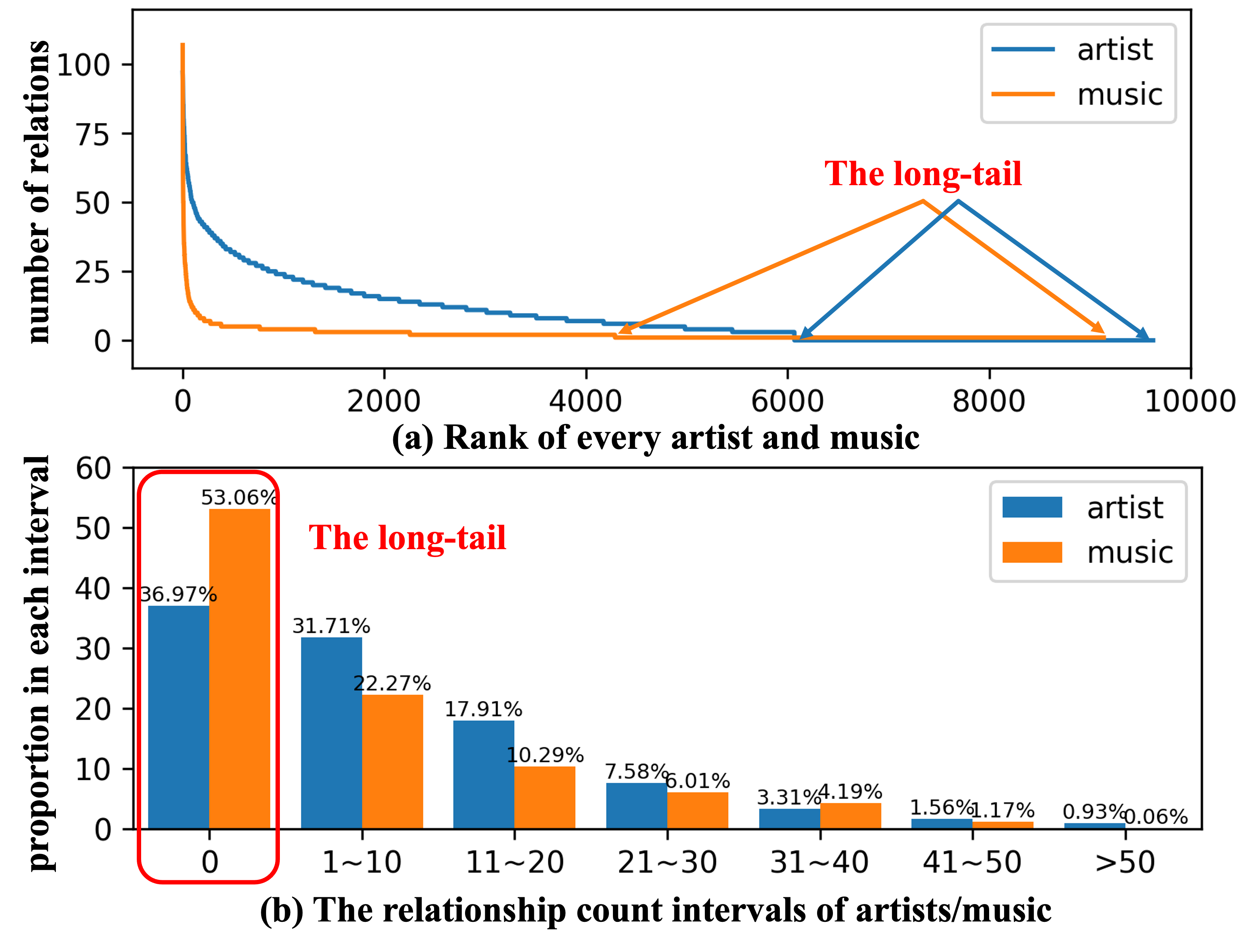}
\caption{The long-tail distribution of artists/music in Allmusic dataset. (a) is the relations number distribution of every artist/music, while (b) divides the relations number into intervals and counts the proportion of artists/music in each interval.}
\label{fig1}
\end{figure}

How to devise a better similarity representation of artist and music is a classic topic in the music information retrieval area \cite{knees2016music, pampalk2006audio}, and similarity-based artist or music recommendation is also widely used in commercial music applications. In previous works, there are two types of information can be utilized to model and calculate similarity: 
\begin{enumerate}
    \item interactive information, \cite{korzeniowski2021artist} defines similarity based on social relationship , while \cite{dhruv2019artist, hansen2020contextual} uses with user-item information, according to \cite{wang2022ssar, salha2021cold}, co-occurrence information is deemed to be useful for modeling similarity;
    \item composition-related information, \cite{deldjoo2021content}, which utilizes music content and meta information to model the artist/music and compare the distance between them in feature space to determine the similarity \cite{lee2020metric, huang2020large, schindler2019multi, thome2022musical, cleveland2020content, fessahaye2019t, lerch2023music}.
\end{enumerate}
These existing approaches have limited focus on addressing the long-tail problem \cite{levy2010music}, despite its common occurrence and the challenges associated with achieving robust performance. We define artists and music without relationship labels as the long-tail category and Figure \ref{fig1} depicts the distribution of these long-tail entities within the Allmusic dataset. The long-tail part of artists accounts for $36.97\%$ and music for $53.06\%$. This paper proposes a Long-Tail Friendly Representation Framework (LTFRF) that considers the factors affecting music and artist representation from multiple perspectives to optimize the weak representation problem of long-tails.

Our approach involves utilizing music content information, music meta-information, and user-interaction information as inputs, and leveraging various relationship information as labels to train a general representation model via metric learning. We subsequently combine this model with the Graph Neural Network (GNN) to measure the similarity of artists and music\cite{wei2019mmgcn, korzeniowski2021artist, salha2021cold}. Notably, we have observed that the general representation model that integrate multiple types of information demonstrate a significant improvement in their generalization ability towards long-tail entities.


\section{Long-Tail Friendly Representation Framework}
\label{sec:LTFRF}

\begin{figure}[htbp]
\centering\includegraphics[scale=0.22,trim=0 0 0 0]{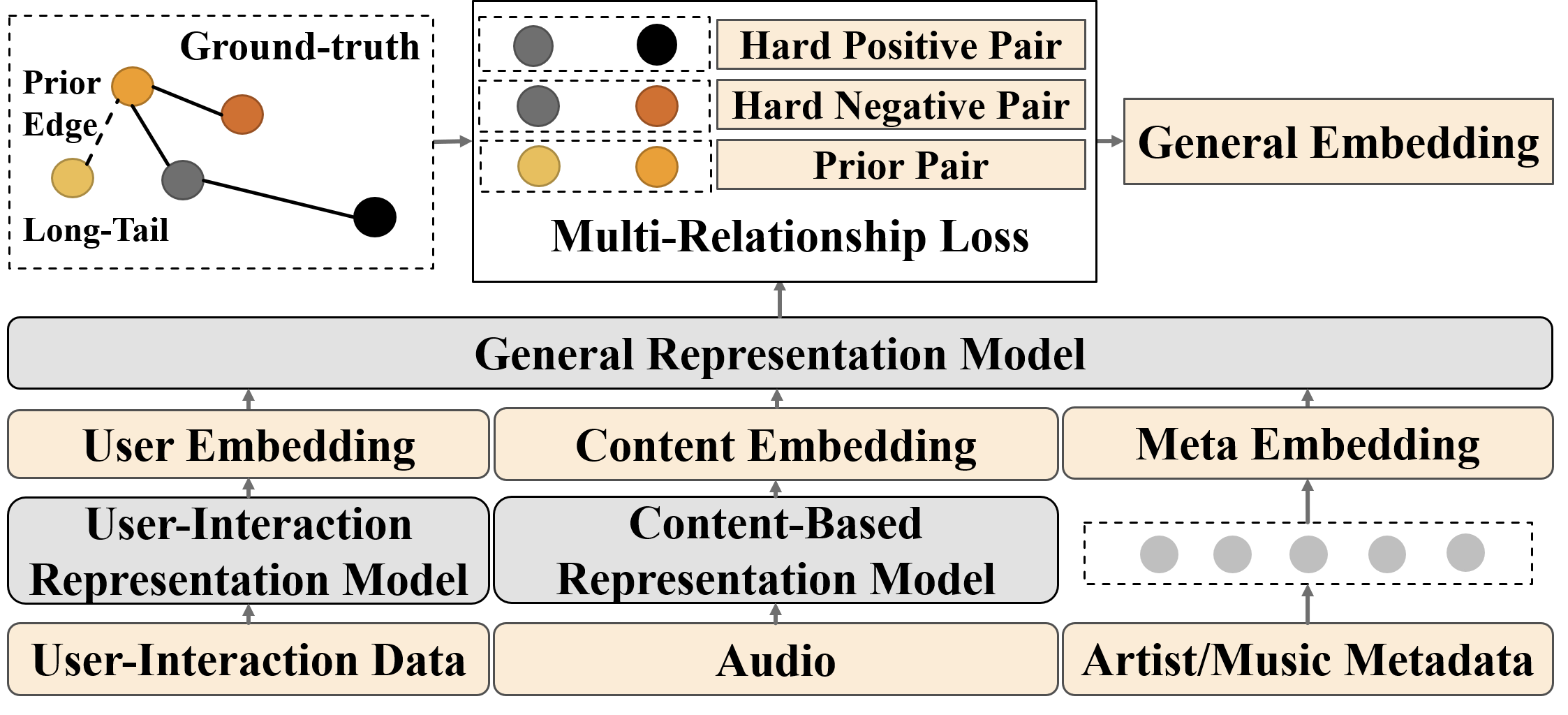}
\caption{The overall architecture of LTFRF. The optional GNN backend is not included.}
\label{fig2}
\end{figure}

The whole framework is shown in Figure \ref{fig2}, and the details of it are shown in Figure \ref{fig3}. Content-based Representation Model (CbRM) \cite{huang2020large, cleveland2020content} takes the music compositions as input, and establishes the relationship between artists/music representation and output content-based embedding $E_C$. User-interaction Representation Model (UiRM) decomposes the similar matrix defined by the user to get user-interaction embedding, $E_U$. The core General Representation Model (GRM) encodes the meta information of artists/music into hidden vectors and combines $E_C$ and $E_U$ as input to obtain general embedding, $E_G$. It is trained through supervised learning with meta-consistency relationships and social relationships among artists/music. 

\subsection{Content-based Representation Model}

CbRM is shown in the lower right in Figure \ref{fig3}. It maps $C_{ik}$, music work $k$ of artist $i$, to music embedding $E_{C,ik}$. We define the similarity relationships $R_C$ of music contents by whether they belong to the same or similar artists. If so, the content-based similarity label of artist $i$ and artist $j$ is the same, $Y_{C,i}=Y_{C,j}$. We conduct semi-supervised metric learning and mine the difficult positive and negative samples. In this way, even long-tail artists/music without similar relationships can be trained without labels, indirectly enhancing the expression of long-tailed artists/music. Finally, we get content-based artist embedding $E_{C,i}$ by aggregating $K$ songs with the highest popularity ($pop_i$): 
\begin{equation}
E_{C,i} = \frac{1}{K}\sum_{C_{ik} \in pop_i} E_{C,ik}.
\end{equation}

For music, $E_C$ is equal to $E_{C,ik}$. The embedding $E_C$ output by CbRM measures the similarity of artists/music in several content-based aspects: genre, mood, timbre, chord, etc.

\subsection{User-interaction Representation Model}
\begin{figure}[!t]
\centering\includegraphics[scale=0.25,trim=0 0 0 0]{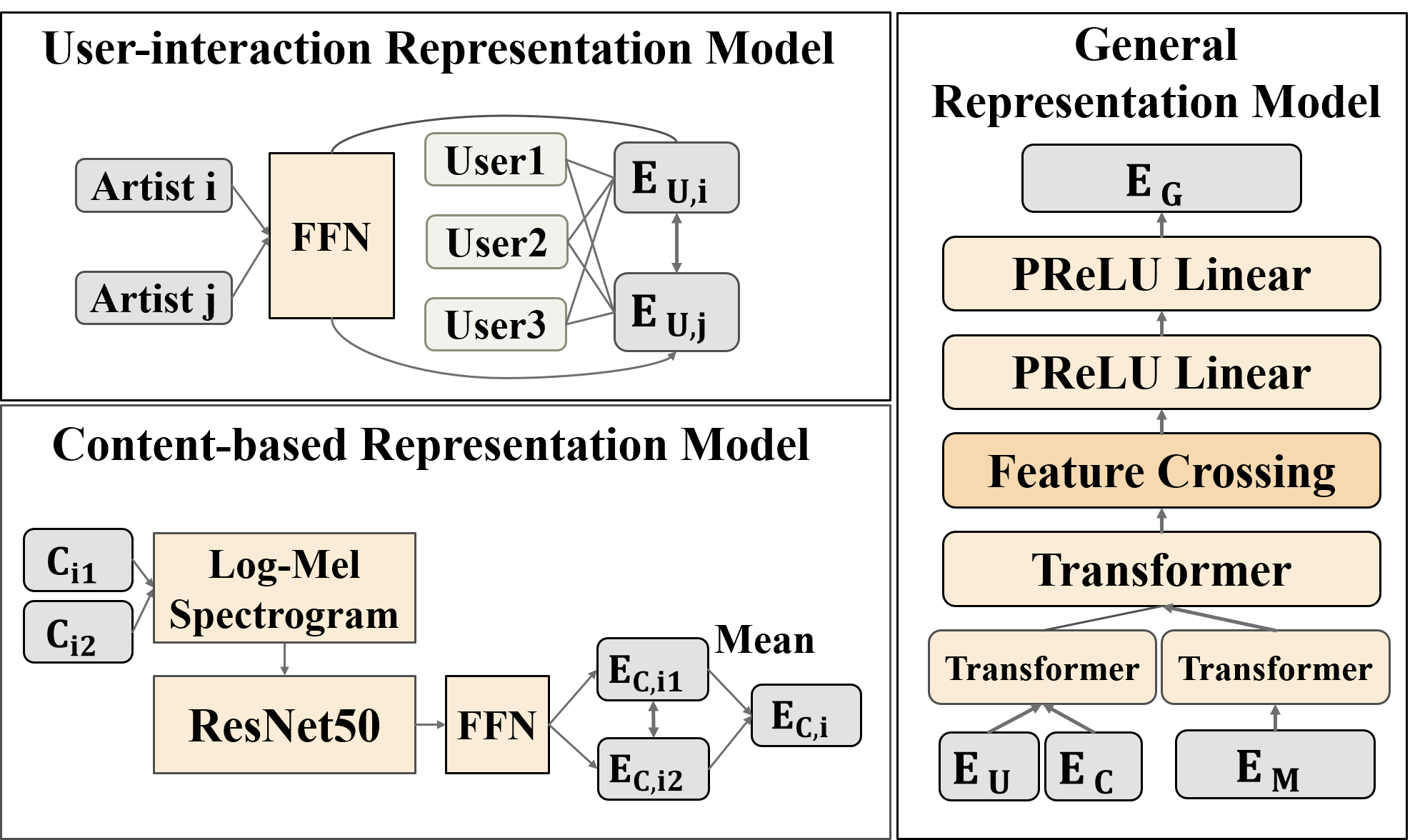}
\caption{Details of each module of LTFRF. CbRM is shown at the lower right, UiRM at the upper right and GRM at the left.}
\label{fig3}
\end{figure}

UiRM extracts the relevant information from the user's behaviour data, as shown at the upper right of Figure \ref{fig3}. UiRM refers to the generalized matrix decomposition method based on a neural network to obtain user-interaction embedding $E_U$ \cite{sunitha2018music,Liang2015content}. We define the User-based similarity relationship $R_U$ through user behaviour. If multiple users interact with item $i$ and item $j$ simultaneously, both items have the same user-defined similarity label $Y_U$: $Y_{U ,i} = Y_{U,j}$, and the item can be an artist or music. We only have artist-related user data in this paper. Through user interaction, some long-tailed artists and famous artists are connected.

$R_U$ can be calculated in a variety of ways. For example, the two items are similar when the number of users shared by item $i$ and item $j$ exceeds a certain threshold. It is also possible to define that the item $i$ is similar to top $K$ items having the most common users. We can also combine the two ways to calculate the similarity score.

\subsection{General Representation Model}
The architecture of GRM is shown at the left of Figure \ref{fig3}, which is the essential part of LTFRF. Suppose $E_M=[E_{genre}, E_{region}, E_{popularity},…]$ is the artist/music metadata embedding. The first two transformer layers fuse $E_C$, $E_U$ and $E_M$ and obtain embedding $E \in \mathbb{R}^{D}$. Then we divide $E$ into $Z$ fields, namely $E^{(1)}=[E_1, E_2,\dots, E_Z]^T \in \mathbb{R}^{Z\times F}$, and use self-attention of the third transformer layer to weight the feature importance of each field to integrate multi-source features.

Then we combine the multiple fields feature through feature crossing, obtain the second-order feature $E_{cf}$, and concatenate it with the original feature $E^{(2)}=[E^{(1)}, E_{cf}]$. The general embedding $E_G$ is finally obtained by multiple Linear layers activated by Parametric Rectified Linear Unit (PReLU).


\subsection{Multi-Relationship Loss}
\label{sec:loss}
As mentioned above, we define the content-based similarity relationship $R_C$ between songs in CbRM and the user interaction relationship $R_U$ in UiRM. GRM uses two relations to determine the similarity. The first is the reliable similarity relationship $R_A$ mined from various sources as the ground truth. The second is the relationship $R_P$ constrained by artist/music metadata and prior knowledge. With similarity metric function $f$, the loss function of all model training in LTFRF can be defined by supervised metric learning \cite{lee2020metric,wu2020effective,won2021multimodal}. 

Let $X\in \{C,U,A\}$, and denote the similarity matrix of $N$ artist/music embeddings $E_{X}$ as $S_X \in \mathbb{R}^{N \times N}$. The label defined by $R_X$ is $Y_X$. $S_{X,ij}=f(E_{X,i}, E_{X,j})$ is the similarity of embedding $E_{X,i}$ and embedding $E_{X,j}$ under the similarity measure $f$. Multi-Similarity Loss \cite{wang2019multi} considers three similarities for pair weighting, providing a more principled approach for collecting and weighting informative pairs. We first combine the similarity matrix $S_{X}$ and relationship label $Y_{X}$ to mine informative pairs. Given a margin $\delta$, the index set of positive pair $(E_{X,i}, E_{X,j})$ is expressed as: 
\begin{equation}
\mathcal{P}_{X,i}=\left\{j\Big|S_{X,ij} < \min\limits_{Y_{X,k} \neq Y_{X,i}} {S_{X,ik} + \delta} \right\}.
\end{equation}
Similarly, the index set of negative pairs $(E_{X,i}, E_{X,j})$ is: 
\begin{equation}
\mathcal{N}_{X,i}=\left\{j\Big|S_{X,ij} > \min\limits_{Y_{X,k}=Y_{X,i}} {S_{X,ik} -\delta} \right\}. 
\end{equation}
Finally, Multi-Similarity Loss is defined as: 
\begin{equation}
\label{msloss}
\begin{aligned}
\mathcal{L}_{X,MS}=\frac{1}{N}\sum_{i=1}^{N}\Big\{ \frac{1}{\alpha_1}log\Big[1+\sum_{j\in \mathcal{P}_{X,i}} e^{-\alpha_1 (S_{X,ij} - \gamma_1)}\Big]\\+ \frac{1}{\beta_1}log\Big[1+\sum_{j\in \mathcal{N}_{X,i}}e^{\beta_1 (S_{X,ij} - \gamma_1)}\Big]\Big\},
\end{aligned}
\end{equation}
where $\alpha_1$, $\beta_1$ and $\gamma_1$ are hyper-parameters.

Multi-Similarity loss unifies the training targets of CbRM, UiRM and GRM and utilizes multi-source relationship information. However, for the long-tail problem, there is a large proportion of long-tail artists/music without similarity edges. The trained long-tail embedding will be less accurate when no relationship information exists. Fortunately, since we have artist/music metadata, we use prior knowledge to construct a meta-consistency relation $R_P$ so that the artist/music representation space is smoother. In GRM, we think similar artists/music will have a similar genre, region and popularity. When item $i$ and item $j$ have the same genre, region and popularity, we set $Y_{P,i}=Y_{P,j}$. The item can be an artist or music. 

The prior loss $\mathcal{L}_{Prior}$ is obtained by replacing $\mathcal{P}_{X,i}$ and $\mathcal{N}_{X,i}$ in the Multi-Similarity Loss with the positive meta-consistency relationship set $\left\{j\big| Y_{P,i}=Y_{P,j}\right\}$, and negative meta-consistency relationship set $\left\{j\big| Y_{P,i}\neq Y_{P,j}\right\}$, respectively. It uses a set of different hyper-parameters $\alpha_2$, $\beta_2$ and $\gamma_2$. The prior loss is a regularization term to smooth the representation space of artist/music. The final Multi-Relationship Loss of GRM is: 
\begin{equation}
\mathcal{L}_{MR} = \mathcal{L}_{A,MS} + \lambda \mathcal{L}_{Prior},
\end{equation}
where $\lambda$ is the hyper-parameter that balances the two losses. When the value of $\lambda$ is appropriate, $\mathcal{L}_{Prior}$ removes the unreliable noise in the mined relationship and has a less negative impact on the real similarity relationship.

\section{Experiments}
\label{sec:Experiments}
\subsection{Experiments Setup}

We build a multi-source dataset, which includes artist/music metadata, artist music data, relationship data and user-interaction data. Only user-artist data exists in the user interaction dataset, but no user-music data exists. The dataset is collected from AllMusic\footnote{https://www.allmusic.com/}.  

The multi-source artist dataset includes $9627$ artists in total. The training, validation and test include $7703$, $962$ and $962$ artists, respectively, and each artist has an average of $3.8$ songs. $36.97\%$ of the artists in the dataset are long-tail artists. The user-artist interaction dataset is built by a totally of $6.8$ million data, and there are $65574$ similarity relationships in the artist relationship data. We also collected a music similarity dataset to verify the validity of LTFRF on music. The dataset includes $61584$ pieces of music with $82880$ music similarity relationships in total. $53.06\%$ of the music in the dataset is long-tail music.

Hit Ratio@K (HR@K), MAP, NDCG, and Consistent@K are used to evaluate the LTFRF performance. The first three metrics will also be used to assess the performance of similar music recommendation tasks. Due to the long-tail having no similarity information, we use Consistent@K as an indirect metric to evaluate the recommendation performance of the long-tail. Consistent@K is defined by calculating the same number of meta tags between two artists or music.

\subsection{Results and Discussion}
\begin{figure}[htbp]
\centering\includegraphics[scale=0.32,trim=0 0 0 0]{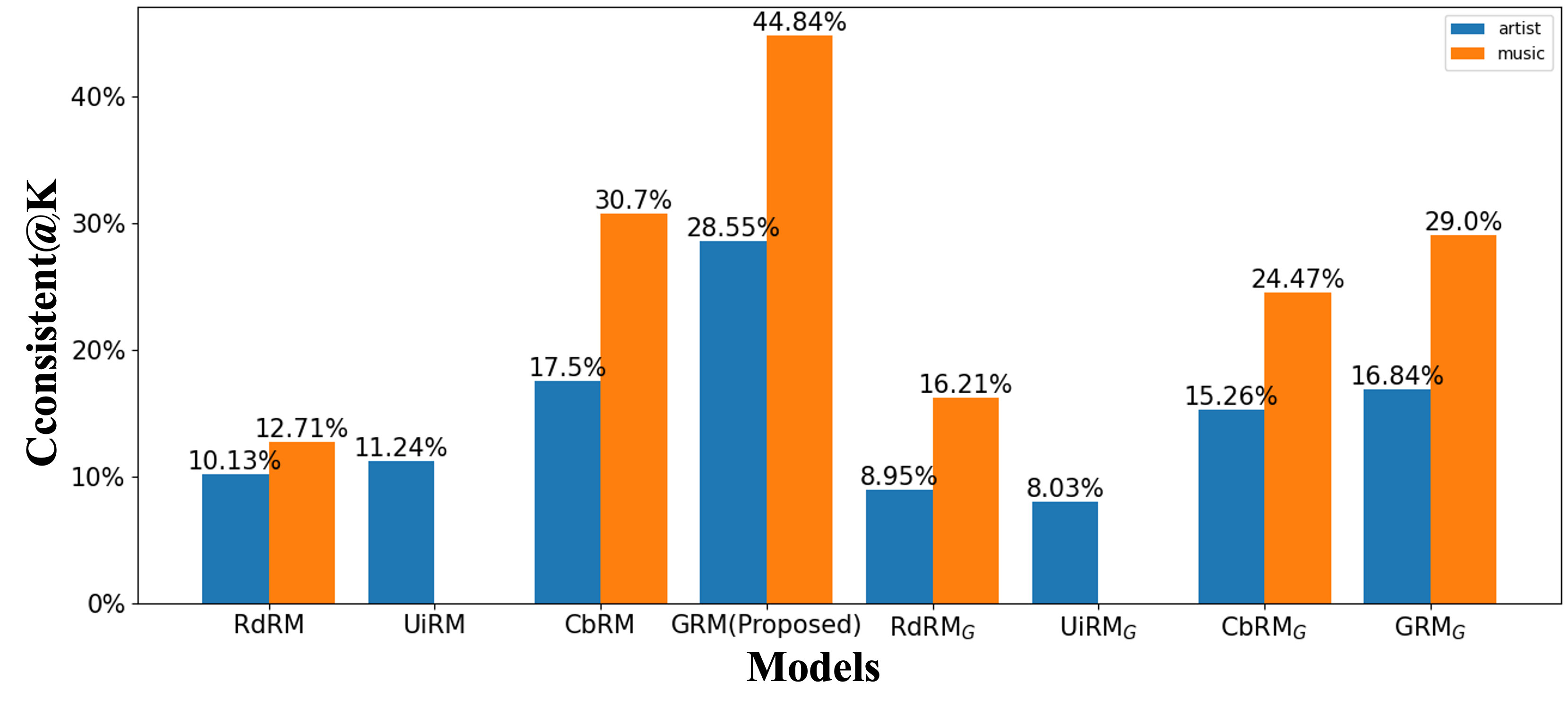}
\caption{Performances of different models in the long-tail test set, the metric is Consistent@10.}
\label{fig5}
\end{figure}

We compared the artist/music recommendation results of different representation models in the experiment. Among them, $CbRM$ is the baseline, which is modified from \cite{huang2020large}. We also use GNN as the backend of these representation models to compare them with LTFRF. Among them, $RdRM_G$ using random embedding as the node embedding of $GNN$ is also a classical method \cite{korzeniowski2021artist}. 

\begin{table}[h]
\caption{Performance with embeddings output by each representation model on the AllMusic test set of the multi-source artist dataset. In Consistent@10, the two results columns represent the performance of all artists / long-tail artists, respectively.}
\label{table1}
\centering\small
\begin{tabularx}{8.5cm}{ccccc}
\toprule
Methods &HR@10 &Consistent@10 &MAP &NDCG \\
\midrule
$RdRM$ &1.47\% &11.23\% / 10.13\% &1.08\% &0.1803 \\
$UiRM$ &4.30\% &12.88\% / 11.24\% &3.83\% &0.2144 \\
$CbRM$ &30.29\% &54.88\% / 17.50\% &17.87\% &0.4080 \\
$GRM$ &\textbf{39.98\%} &\textbf{86.19\% / 28.55\%} &25.68\% &0.4846 \\
\bottomrule
$RdRM_{G}$ &21.90\% &28.31\% / 8.95\% &14.77\% &0.3553 \\
$UiRM_{G}$ &26.91\% &42.67\% / 8.03\% &18.71\% &0.4045 \\
$CbRM_{G}$ &34.76\% &57.04\% / 15.26\% &23.60\% &0.4618 \\
$GRM_{G}$ &39.89\% &65.03\% / 16.84\% &\textbf{26.14\%} &\textbf{0.4904} \\
\specialrule{0.05em}{2pt}{2pt}
\end{tabularx}
\end{table}
\normalsize

Table \ref{table1} shows the performance of representation models mentioned in the paper. $X(y)RM_G$ means we regard embedding $E_X$ as the node feature of the artist similarity graph and use $GNN$ to train the artist similarity further, for example when $Xy = Cb$, that means we use the embedding $E_C$ generated by $CbRM$ as node feature to further train $GNN$.

From the experimental results, we find the performance of $GRM$ on famous artists is significantly better than $RdRM$, $CbRM$ and $UiRM$, whether or not to use $GNN$ as the backend model. It shows that artists' music works and metadata are essential information for measuring the similarity of artists, and the user interaction relationships also provide a positive impact. From the perspective of Consistent@K, the addition of prior knowledge significantly improves the representation ability of long-tailed artists. Artists with the same meta tags can naturally gather together in the artist representation space, and the possibility of similarity between these artists will be significantly improved. 


After joining $GNN$ training, there are no significant improvements in every metric, and the Consistent@K result even decreased when using $GRM$ as the frontend model. Because $GRM$ has fully integrated various information related to artists, $GNN$ can no longer provide more helpful information. Besides, $GNN$ highly relies on relationship data as a graph edge during training, while $GRM$ does not completely require it, leading to $GRM$ being better than $GRM_G$ in representing long-tail artists.


\begin{table}[h]
\caption{Performance with different embeddings on the test set of the music similarity dataset.}
\label{table2}
\centering\small
\begin{tabularx}{8.5cm}{ccccc}
\toprule
Methods &HR@10 &Consistent@10 &MAP &NDCG \\
\midrule
$RdRM$ &0.34\% &12.95\% / 12.71\% &0.18\% &0.1425 \\
$CbRM$ &23.23\% &33.65\% / 30.70\% &13.52\% &0.3686 \\
$GRM$ &42.65\% &\textbf{45.44\% / 44.84\%} &27.23\% &0.5220 \\
\bottomrule
$RdRM_{G}$ &25.26\% &17.78\% / 19.20\% &16.21\% & 0.3825\\
$CbRM_{G}$ &37.05\% &34.96\% / 30.34\% &24.47\% & 0.4891\\
$GRM_{G}$ &\textbf{44.58\%} &37.82\% / 34.03\% &\textbf{29.00\%} & \textbf{0.5476}\\
\specialrule{0.05em}{2pt}{2pt}
\end{tabularx}
\end{table}
\normalsize



The representation advantages of LTFRF can also be reflected in similar music recommendation tasks. As shown in Table \ref{table2}, $RdRM$ is random music embedding. We still use $CbRM$ as the music recommendation baseline and get the music tagging embedding \cite{schindler2019multi} from it. Due to the lack of user-music interaction data, we do not apply the UiRM in similar music tasks so that the $GRM$ is also trained without user-interaction data. Nevertheless, the performance of $GRM$ in similar music recommendation tasks has also been significantly improved. It indicates that $E_G$ can represent artists, music, and relationship information well. Consistent with the conclusion of previous experiments, there is no significant improvement in all metrics after joining GNN training. 

In general, the proposed model achieves 9.69\%/19.42\% better than the baseline in Hit Ratio@10 and the performance of all models on the long-tail is shown in Figure \ref{fig5}. The proposed model is significantly better than the baseline $CbRM$, which is 11.05\% higher on the artist task and 14.14\% higher on the music task.
Besides, LTFRF can also be used in other similarity modeling tasks.

\section{Conclusion}
\label{sec:Conclusion}
In this paper, we propose a Long-Tail Friendly Representation Framework based on metric learning and provide a unified framework to model similar relationships such as artist similarity and music similarity. The Multi-Relationship Loss combined with prior knowledge alleviates the long-tail problem. It improves the poor expression ability of the long-tail and makes their representation more accurate. Moreover, it achieves the best results on similar artist and music recommendation tasks. The numerical results and analyses in experiments verify that LTFRF can obtain general embedding with good generalization and robustness.




\bibliographystyle{IEEEbib}
\bibliography{IEEEbib.bib}

\end{document}